\documentclass[useAMS]{mn2e}

\usepackage{graphicx}

\begin{document}

\title[Massive Star Forming Regions IRAS 21413+5442 and IRAS 21407+5441]
{Near-Infrared Photometry and Radio Continuum Study of the Massive Star Forming Regions
IRAS 21413+5442 and IRAS 21407+5441}
\author[B.G. Anandarao et al]{B.G. Anandarao${^1}$
\thanks{anand@prl.res.in}, V. Venkata
Raman${^1}$\thanks{vvenkat@prl.res.in}, S.K. Ghosh${^2}$
\thanks{swarna@tifr.res.in},
\newauthor D.K. Ojha${^2}$
\thanks{ojha@tifr.res.in} and M.S.N. Kumar${^3}$
\thanks{nanda@astro.up.pt}  \\
$^1$Physical Research Laboratory (PRL),
Navrangpura, Ahmedabad - 380009, India \\
${^2}$Tata Institute of Fundamental Research (TIFR),
Homi Bhabha Road, Mumbai - 400005, India \\
${^3}$Centro de Astrofisico da Uninversidad do Porto,
Porto, Portugal}

\date{}

\pagerange{\pageref{firstpage}--\pageref{lastpage}} \pubyear{2007}

\maketitle

\label{firstpage}

\begin{abstract}
IRAS 21413+5442 and IRAS 21407+5441 are two massive star forming regions
of high luminosity, likely associated with each other.
Near-infrared photometry on these two IRAS sources was
performed at UKIRT using the UFTI under excellent seeing
conditions yielding an angular resolution of $\sim$ 0.5 arcsec.  
Our results reveal details of stellar content to a 
completeness limit (90\%) of J = 18.5, H = 18.0, and K = 17.5 mag in the
two regions. In IRAS 21413+5442, we
identify a late O type star, having large (H-K) color, to be near the centre of the CO jets
observed by earlier authors. The UKIRT images reveal in  
IRAS 21407+5441, a faint but clear compact HII region around a central 
high - intermediate mass star cluster. 
We have detected a number of sources with large (H-K) color which are not detected in J band. 
We also present the GMRT radio continuum map at 1.28 GHz covering 
the entire region surrounding the two star forming clouds.
The radio continuum fluxes are used to estimate the properties of HII regions 
which seem to support our near-IR photometric results.  
Based on our radio continuum map  
and the archival MSX 8.2 $\mu$m image, we show that the two IRAS sources likely 
belong to the same parent molecular cloud and conjecture that 
a high mass star of large IR colors, present in between the two sources, 
might have triggered star formation in this region. 
However one can not rule out the alternative 
possibility that Star A could be a nearby foreground star. 
\end{abstract}


\section{Introduction}
Appearance of a compact (CHII) or an ultra-compact HII (UCHII) region embedded in a
molecular cloud signifies the formation of a massive star of
spectral type earlier than $\sim$ B3 (Shepherd \& Churchwell
\cite{shep96}; Churchwell \cite{chur02}). During this phase the massive star is believed
to be close to its zero-age main-sequence (ZAMS), although pre-main-sequence 
manifestations such as outflows have been detected in some
cases (e.g., Weintraub \& Kastner \cite{wein96}; Beuther et al. \cite{beut02};
Kumar et al. \cite{kuma06}). 
Massive star formation occurs always in clusters. Further 
the environment around the massive stars gets affected by their 
strong winds and energetic radiation.
It is therefore important to study such CHII/UCHII
regions in order to find out possible evolutionary stage of
stellar content in their vicinity. Near-Infrared photometry was shown to be very
useful for this purpose (e.g., Lada \cite{lada85}). Supplementary radio
continuum measurements can provide important physical
parameters concerning the object under certain assumptions
(e.g., Scheffler \& Elsasser \cite{sche88} and references therein).
In this paper we describe a study in near-IR and radio continuum in and around two 
IRAS sources that are believed to be massive star forming regions by  
virtue of the presence of CHII/UCHII.

IRAS 21413+5442 (Object 1) is one of the highly luminous, massive
young stellar objects (YSOs) in our Galaxy and is situated at an 
estimated distance
of 7.4 kpc (Wouterloot \& Brand \cite{wout89}; Yang et al \cite{yang02}). 
This source is
identified with the presence of a UCHII region, called IRAS 21413+5442S,
about 20 arcsec south of a compact HII region, 
called IRAS 21413+5442N (Miralles et al.
\cite{mira94}). The far infrared luminosity from
IRAS fluxes at 12, 25 and 60 $\mu$m was estimated to be 3.2 $\times$ 10$^5$
L$_\odot$ (Campbell et al.
\cite{camp89}). The CO surveys of Shepherd \& Churchwell
\cite{shep96} have classified this source as a massive star
forming region with outflows of high velocity gas. This
indicates therefore the pre-main-sequence signature of the
source that is believed to be at/near its ZAMS stage (due to the
appearance of a UCHII region). 
Bronfman et al. \cite{bron96} detected this object in their 
survey of CS(2-1) rotational
emission (at 97.981 GHz) that is believed to be a signature of
high density molecular gas.
Ishii et al.
\cite{ishi98,ishi02} have identified the 3.1 $\mu$m H$_2$O
ice in absorption which indicates the presence of a disk or
thick nebular matter around the object. Ishii et al.
\cite{ishi98,ishi02} classify this object as heavily obscured
due to the larger excess in near-IR colors. Interestingly,
Shepherd \& Churchwell \cite{shep96} have clearly mentioned
that the source engine of the outflows (or jets) found in this
object must be a heavily obscured (pre-main-sequence) star yet
to be discovered. 

The second object in our study namely IRAS 21407-5441 (Object  
2) was classified as a UCHII region from its IRAS colors and high 
luminosity of 8 $\times$ 10$^{4}$ L$_{\odot}$ (Wood
\& Churchwell \cite{wood89}: S$_{25}$/S$_{12}$ $\geq$ 3.7 and
S$_{60}/S_{12}$ $\geq$ 19.3). The object however fell short of
the sensitivity limit of Bronfman et al. \cite{bron96} survey of CS(2-1) 
line, possibly due to its low density gas.
This object is situated at about the same (kinematic) distance ($\sim$ 8 kpc) as Object 1, 
and is in all likelihood associated with it 
(as quoted in Carral et al \cite{carr99} based on the 
rotation curve due to Wouterloot et al \cite{wout90}).
At this distance, the two Objects are separated in projection by about 20 pc (9.5 arcmin).
The MSX A band (8.2 $\mu$m) image of the region seems to support the 
association of the two IRAS sources; it also shows an interesting bright point source 
nearly mid-way between the two IRAS sources (called Star A in Fig 1 
and subsequent sections of the paper).

In this paper we describe the first sub-arcsec UKIRT 
near-infrared photometry and 1.28 GHz radio continuum mapping 
in an attempt to classify the stellar 
content in and around the two IRAS sources and to try and establish their association.
We also examine the nature of Star A and its possible role in the region. 
We have also made use of archival 2MASS and MSX data to supplement our study.   
In section 2.1, the near-infrared photometry is presented and 
in section 2.2, we present 1.28 GHz radio continuum mapping of the two objects. In section 3, we 
describe the results from near-infrared photometry and radio continuum observations. 
Section 4 gives discussion of the results that reveal some new 
interesting insights on the two objects. Section 5 lists important conclusions of the 
present work. 

\section {Observations}

Fig 1, generated from the 2MASS archival images, shows the two regions of our interest 
namely IRAS 21413+5442 (to the left, Object 1) and IRAS 21407+5441 (to the right, Object 2). 
Marked in the figure are  
three clusters - one in IRAS 21413+5442 (Cluster 1) and two in IRAS 21407+5441 (Clusters 2C and 2N). 
A bright star with large visual extinction, marked as Star A, 
is present in between the two IRAS sources. Also shown in the figure are the MSX low resolution 
contours in A band (8.2 $\mu$m) overlaid on the 2MASS image. One may notice the apparent 
association of the two IRAS sources with the Star A nearly at the middle.   
The observations made on these sources are described in the following 
two subsections.

\subsection{UKIRT-UFTI Photometry at sub-arcsec resolution}
Sub-arcsec resolution JHK photometry was performed at UKIRT
during 17 July 2005 on the two objects under excellent seeing
conditions ($\sim$ 0.5 arcsec) using the UFTI (HAWAII-1 1024
$\times$ 1024) camera. The overall integration times were 75
sec for each of the three bands. The plate scale is 0.09
arcsec/pixel that gives a useful central field of view of $\sim$
75$^{\prime\prime}$ on the two objects. The standard star used
was FS 151 (Hawarden et al. \cite{hawa01}) for both the
sources.

  \begin{figure*}
   \centering
\includegraphics[scale=0.75]{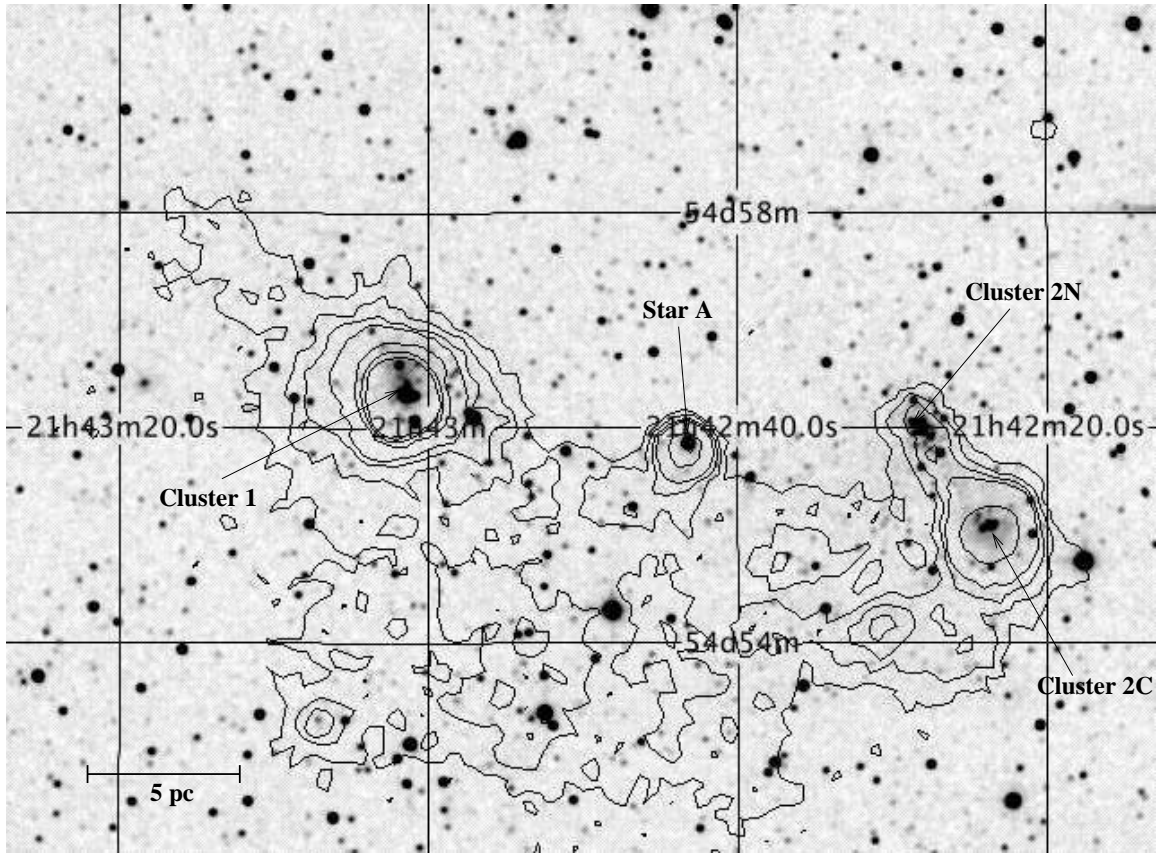}
   \caption{2MASS Ks-band image of the combined field of the Clusters 1 and 2. 
     For Cluster 2 the central source is marked as Cluster 2C and the north-eastern 
      source as Cluster 2N. At the middle of the image, the highly reddened object 
      Star A is shown. The abscissa (RA) and ordinate (Dec) are for J2000 epoch. The 
      overlaid contours represent MSX low resolution image of the region; with 
      contour levels 2, 3, 4, 5, 10, 15 and 20 W/m$^{2}$/Sr. The linear size in the region 
      shown at the bottom left assumes a distance of 7.4 kpc.}
              \label{2mass12}%
    \end{figure*}

  \begin{figure*}
   \centering
\includegraphics[scale=0.75]{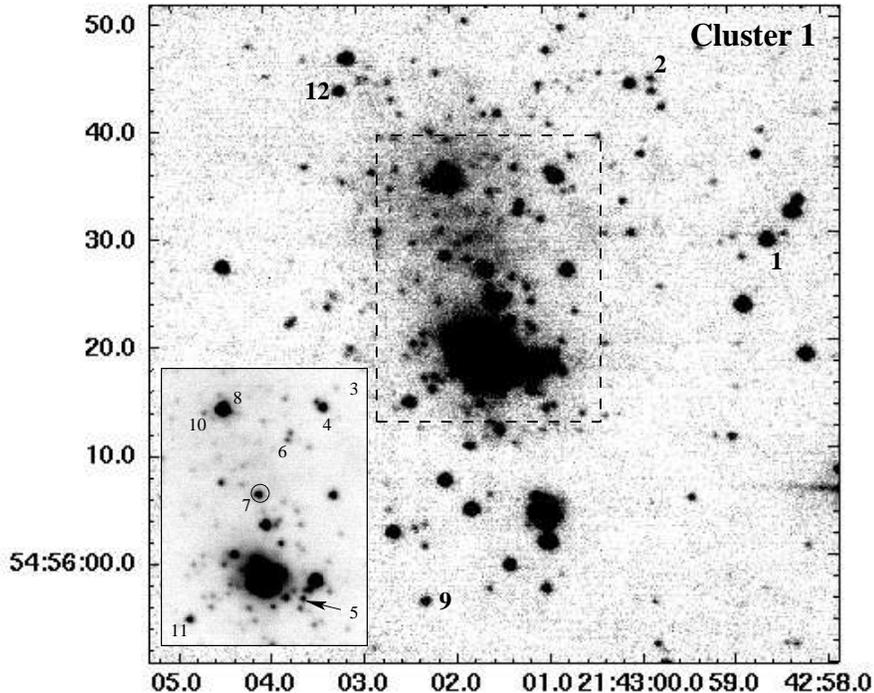}
   \caption{UKIRT K-band image of IRAS 21413+5442 (Cluster 1). The inset at the 
   bottom-left represents the central region (see dashed lines in the main fig) and shows  
   the identified stars with large IR excess falling in the T Tauri star region 
   and beyond (see text and Table 1). The star (No. 7) that was shown encircled is situated at the centre of 
   the CO jets. There are several "red 
    objects" that are detected only in H and K bands with large (H-K) color (see text, Fig 5 and Table 1).
    The abscissa (RA) and ordinate (Dec) are for J2000 epoch.}
              \label{ukirkob1}%
    \end{figure*}

  \begin{figure*}
   \centering
\includegraphics[scale=0.75]{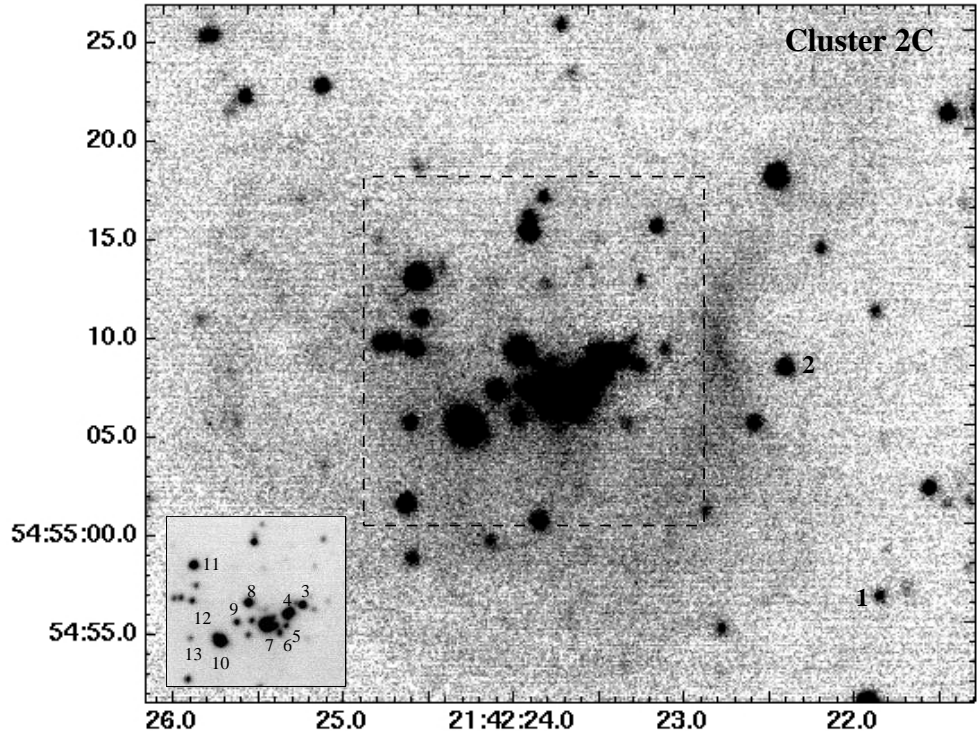}
   \caption{UKIRT K-band image of IRAS 21407+5441:
   Central region (Cluster 2C) revealing the 
   HII region with filamentary structures. The inset at the bottom-left shows 
   stars identified, 
   as in Fig 2 (see also text and Table 2). There are several "red 
    objects" that are detected only in H and K bands with large (H-K) color (see text, Fig 5 and Table 2). 
    The abscissa (RA) and ordinate (Dec) are for J2000 epoch.}
              \label{ukirkob2c}%
    \end{figure*}

 \begin{figure*}
  \centering
\includegraphics[scale=0.75]{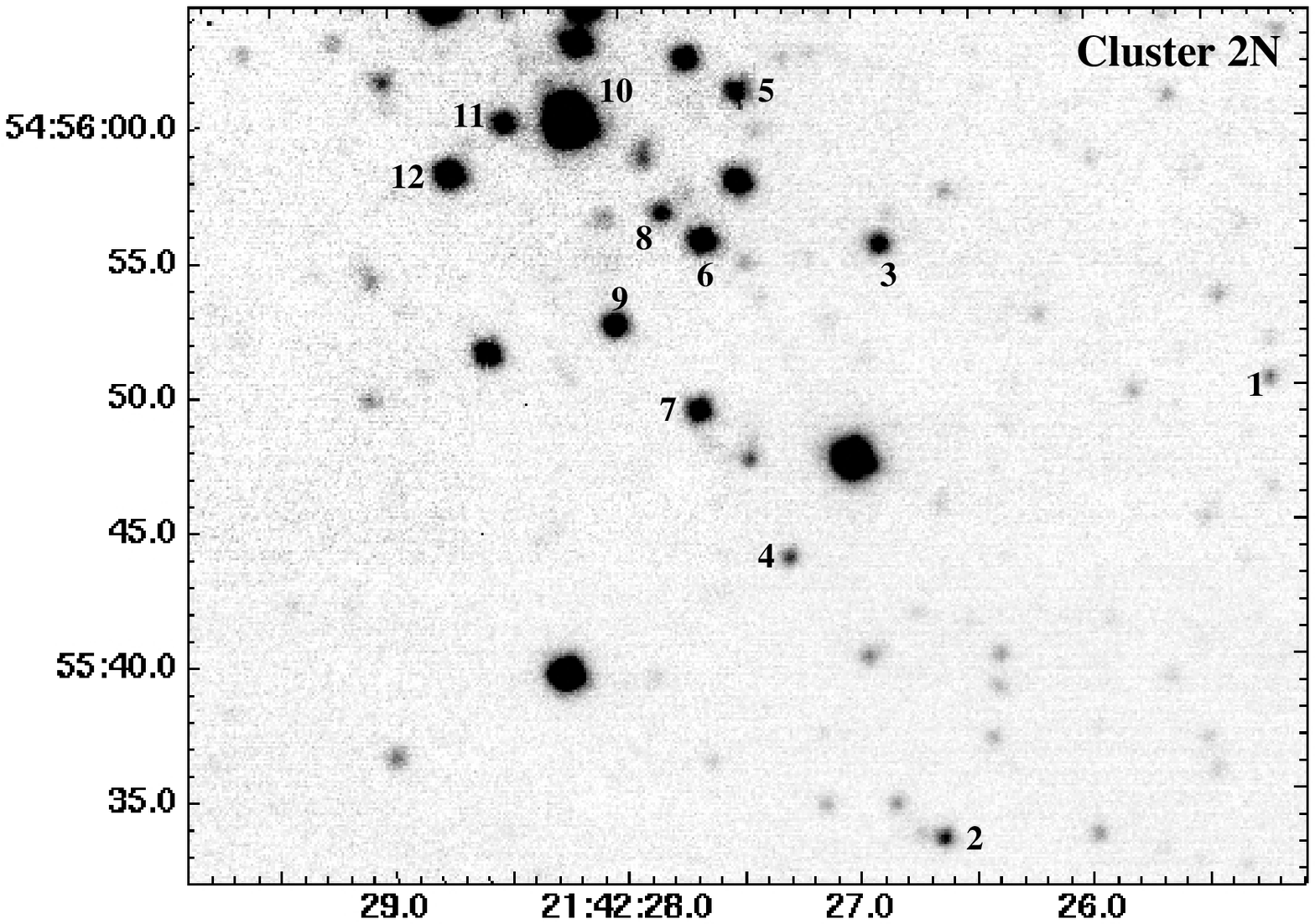}
  \caption{UKIRT K-band image of IRAS 21407+5441: North-eastern region (Cluster 2N). 
   Stars identified are numbered (see text and Table 3). There are several "red 
    objects" that are detected only in H and K bands with large (H-K) color (see text, Fig 5 and Table 3).
   The abscissa (RA) and ordinate (Dec) are for J2000 epoch.}
            \label{ukirkob2n}%
   \end{figure*}
    
The image processing was done by using standard IRAF tasks. Only those sources are
considered which have photometric errors less than 0.1 mag for all the three bands. The
90\% completeness limits are J = 18.5; H = 18.0 and K = 17.5 mag for the
observations on the two objects. 

We found from the MSX image archives (low resolution mosaic images 
at an angular resolution of 18$^{\prime\prime}$, see Price et al. \cite{pric01}), that 
the two IRAS sources in question 
seem to be physically linked, being part of a region that encompasses a large 
sky area (about $8^\prime \times 8^\prime$). Since the UKIRT photometry 
does not cover the entire region of MSX image data for the region, 
we have extracted 2MASS JHK$_{s}$ photometric data for regions in between 
the two IRAS sources, including Star A and its environs.
    
\subsection{1.28 GHz Radio Continuum Observations at GMRT}

The radio continuum observations in the 1.280 GHz frequency band
(with a bandwidth of 16 MHz) were made on 1 August 2003 
at the Giant Metre-Wave Radio
Telescope (GMRT), operated by National Centre for Radio Astrophysics (NCRA), 
Tata Institute of Fundamental
Research (TIFR) near Pune (details of the telescope can be found 
in Swarup et al. \cite{swar91}). Out of the total 30 antennae, 26 had   
operated on the day of observations. The calibration sources used were
3C286 and 3C48 for flux and 2202+422 for phase. The field of
view includes both the objects encompassing a region of
10$^\prime \times 10^\prime$. The image processing was done by
using AIPS software. The data sets were checked for dead antennas, 
bad baseline, bad time ranges,
spikes etc using the tasks UVPLT and VPLOT. The tasks UVFLG and TVFLG 
were used for subsequent editing. Maps of the field were generated 
by Fourier transform and subsequent cleaning and deconvolution using the IMAGR task.
The final angular resolution on the map is $\sim$ 30 arcsec (beam size), after using 
adequate UV tapering on the available base-lines.
The rms noise in the radio map is 1.4 mJy/beam.

 \begin{figure*}
   \centering
\includegraphics[scale=0.7,angle=-90]{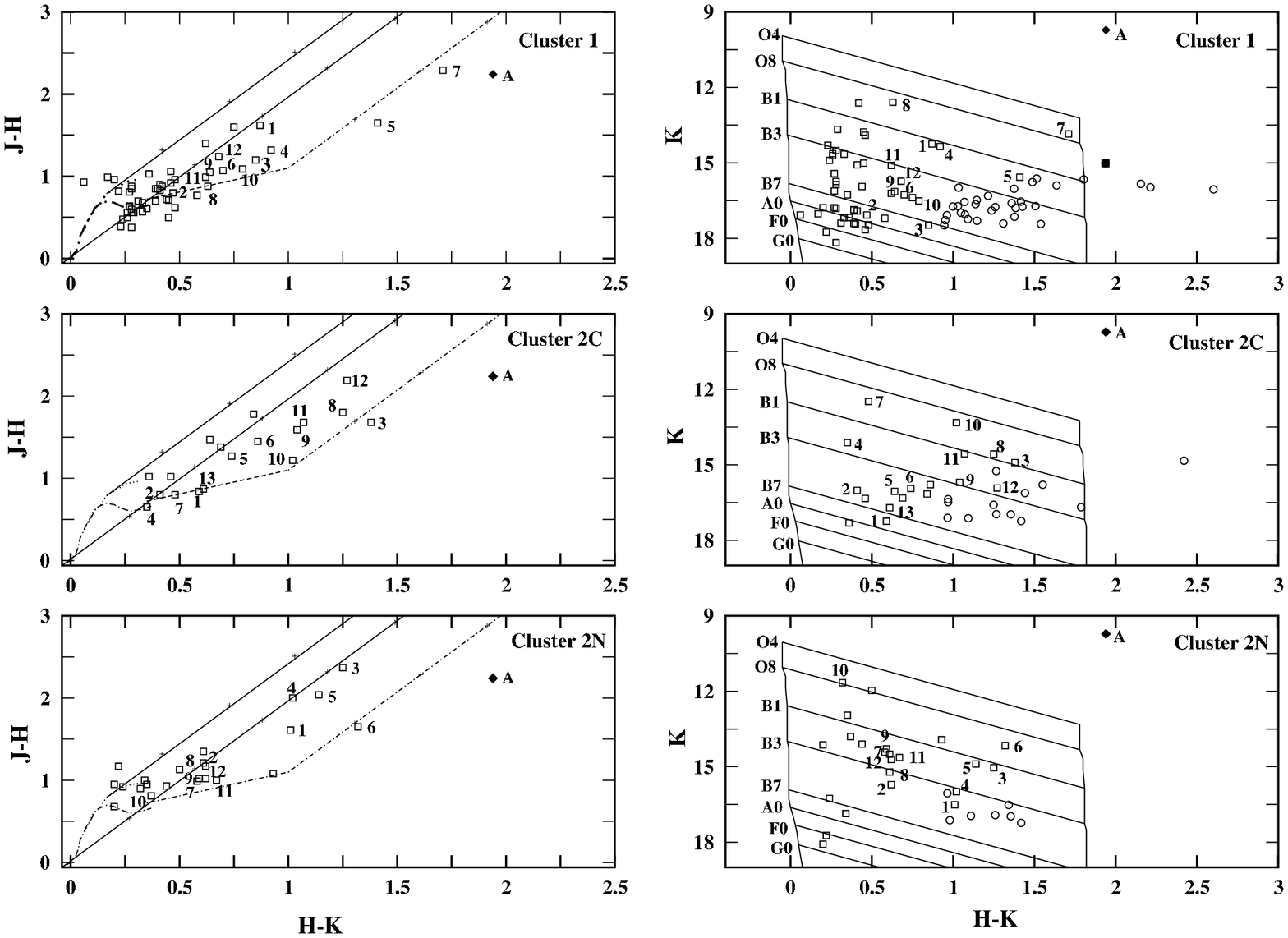} 
   \caption{Color-Color diagrams (left) for the sources in the central HII regions in IRAS 21413+5442
    (Cluster 1, in top panel) and Cluster 2C (middle panel) and Cluster 2N (bottom panel) 
    in IRAS 21407+5441. The thick dashed curve represents the unreddened main-sequence stars; the thin 
    dotted curve shows the unreddened giant stars. The straight lines indicate the extinction 
    vectors for A$_{V}$ = 30 mag and the crosses on the vectors indicate the visual extinction 
    for 5 mag intervals. The T Tauri stars fall on the dashed straight line, while 
    the dot-dashed straight line represents the extinction vector for the T Tauri stars. 
    Color-Magnitude diagrams (rightside panels) are also shown. The vertical lines indicate the unreddened 
    main-sequence stars of different 
    spectral types (on the left-side) and the same with an extinction 
    of A$_{V}$ = 30 mag (on the right-side); 
    the slanting lines showing the extinction vectors. The stars that are common for all the 
    JHK bands are shown by asterisks and the stars that are detected only in HK bands are shown by
    open circles. The filled square represents the embedded star shown circled in Fig 2. 
    The diamond symbol marked as 'A' represents the position of Star A, shown for comparison.}
              \label{i1ccmd}%
    \end{figure*}
    
    

\section{Results}

Here we essentially deal with the three clusters seen in Fig 1: the one 
around the central HII region in IRAS 21413+5442 (Cluster 1); and the central and 
the north-eastern clusters 
in IRAS 21407+5441 (Cluster 2C and Cluster 2N respectively).   
Figs. 2, 3 and 4  show the UKIRT K band images of the
three clusters respectively. The images represent
approximately 90$^ {\prime\prime} \times $90$^ {\prime\prime}$
FOV, although the best image quality and high signal to noise ratio
were restricted to the central region of FOV 75$^ {\prime\prime} \times 75^
{\prime\prime}$. The visual extinction, A$_V$, for each of 
the identified stars was computed from the HK photometric magnitudes, 
using the Bessell \& Brett
\cite{bess88} formulae based on the extinction law of Rieke
\& Labofsky \cite{riek85}, 
with the ratio of total-to-selective extinction $R_{V}$ = 3.1. 
The standard color-color (C-CD)
and color-magnitude (C-MD) diagrams were constructed and shown
in Fig 5 for the three clusters. 
Also shown in the C-CDs are un-reddened
main-sequence locus, un-reddened locus of giant stars (from
Bessell \& Brett \cite{bess88}). The extinction vectors are
shown for B0 and M8 main-sequence stars. The T Tauri star locus
is taken from Meyer et al. \cite{meye97}. 
The C-MD was constructed assuming a distance of 7.4 kpc
for both the sources. The theoretical positions of several
main sequence spectral types are also shown in the diagram with
corresponding reddening vectors. We
have listed in Tables 1, 2 and 3, the positions, H magnitudes, (H-K) and (J-H) colors 
and the visual extinctions (A$_V$) for all the stars detected to the completeness limits in Cluster 1, 
Cluster 2C and 2N respectively. The stars with large
excess falling in the T Tauri region and beyond are shown in serial numbers in brackets. 
These are also marked on the images in Figs. 2, 3 and 4. 
The spectral types obtained from the K vs H-K plots were 
verified against J vs J-H plots and found that there is a very 
good agreement. Considering the 
uncertainties in the exact distance of the sources and in the photometry, 
it is difficult to determine accurate spectral types based only on CM-Ds.
Further the spectral classes would have been slightly over-estimated (by a couple of sub-classes) for 
some sources that have substantial intrinsic infrared excess as they probably 
represent relatively early pre-main-sequence (PMS) stars (Class I).   

We find a number of stars detected only in H \& K-bands but not in J-band
in the UKIRT photometry. These are also shown in the C-MDs in  Fig 5. 
Some of these have large
H-K color excesses ($\geq$ 1.5), especially those that are found in Clusters 1 and 2C.  
These sources could be low/intermediate mass pre-main-sequence stars.
These stars, with (H-K) $\geq$ 1.0, are also listed in Tables 1-3. 

The radio continuum contour map in 1.280 GHz as shown in Fig 6 captures
the two objects IRAS 21413+5442 and IRAS 21407+5441 in a single
observation due to the large field of view ($10^{\prime} \times 10^{\prime}$) of GMRT. 
Also shown overlaid on the radio contours in Fig 6 is the MSX low-resolution mosaic image. 
The figure shows
two dense regions, one around the northern and southern HII
regions in the Object 1 and the other, a more dense and intense
one, around the central cluster of the Object 2 (Cluster 2C; see Fig 6),
extending to encompass the north-eastern cluster (Cluster 2N in Fig 6). 
The radio contours show a maximum around Star A that lies mid-way between 
Object 1 and Objects 2C and 2N. Our radio map compares very well with 
the NVSS 21 cm image of the region. The MSX A band image also correlates well 
with the radio map; the low level extended emission probably representing warm gas/dust.
Fig 7 shows the 1.28 GHz flux density contours overlaid on the 
K band images of Objects 1 and 2 (respectively A and B in Fig 7). 
Since the beam size is about 30 arcsec, the radio map shows the two HII
regions of Object 1 as a single region, with a detected extension consistent with 
the location of these HII regions.

\begin{table*}
 \centering
  \caption{UKIRT JHK photometry of sources in the central HII regions (Cluster 1) of IRAS 21413+5442. 
  The numbers in brackets in the A$_V$ column identify stars with IR excesses in T Tauri region and beyond. 
  The stars listed after the gap are detected only in HK bands.}
  \begin{tabular}{cccccc||cccccc}
  \hline
      R.A.(2000) & Dec.(2000) & H  & [H-K]  & [J-H] & A$_V$ & R.A.(2000) & Dec.(2000) & H  & [H-K]  & [J-H] & A$_V$ \\
\hline
      21:42:58.54 &  54:56:29.61 &   15.11 & 	 0.87 &  1.62 &  14.4 (1)   &  21:43:03.42 &  54:56:16.56 &   17.53 &     0.36 &  1.03 & 5.9  \\ 
      21:42:58.63 &  54:56:39.78 &   17.94 & 	 0.48 &  0.96 &  7.9	&  21:43:03.56 &  54:56:22.51 &    17.7 &     0.31 &  0.70 & 11.6  
      \\ 
      21:42:58.67 &  54:56:37.60 &   16.62 & 	 0.35 &  0.61 & 5.8	&  21:43:03.60 &  54:56:22.08 &   17.51 &     0.33 &  0.68 & 5.4\\
      21:42:58.78 &  54:56:23.60 &   14.37 & 	 0.46 &  0.92 & 7.6	&  21:43:03.75 &  54:56:54.14 &   14.22 &     0.45 &  0.71 & 7.4\\
      21:42:58.80 &  54:56:28.08 &   17.79 & 	 0.39 &  0.70 & 6.4	&  21:43:04.06 &  54:56:48.02 &   17.97 &     0.22 &  0.82 & 3.6\\     
      21:42:58.84 &  54:56:52.38 &   17.32 & 	 0.41 &  0.83 & 6.8	&  21:43:04.30 &  54:56:27.44 &   14.98 &     0.33 &  0.57 & 5.4\\     
      21:42:58.90 &  54:56:29.93 &   18.12 & 	 0.46 &  1.06 &  7.6	&  21:43:04.58 &  54:55:58.98 &   17.05 &     0.27 &  0.81 & 4.5\\     
      21:42:59.30 &  54:56:05.79 &   17.25 & 	 0.39 &  0.85 & 6.4	&  	       &	      & 	&	   &	   &  \\
      21:42:59.78 &  54:56:44.69 &   17.54 & 	 0.47 &  0.80 &  7.8 (2)   &  	       &	      & 	&	   &	   &  \\        
      21:42:59.89 &  54:56:37.68 &   17.08 & 	 0.28 &  0.85 & 4.6	&  21:42:58.88 &  54:56:11.20 &  18.02 &   1.26  &	 & 20.8 \\
      21:43:00.00 &  54:56:44.20 &    15.7 & 	 0.27 &  0.63 & 4.5	&  21:42:59.93 &  54:56:30.27 &  18.23 &   1.51  &	   & 24.9 \\
      21:43:00.07 &  54:56:33.30 &   17.13 & 	 0.06 &  0.93 & 1.0	&  21:43:00.68 &  54:56:17.42 &  17.62 &   1.15  &	   &  18.9 \\
      21:43:00.24 &  54:56:13.41 &   18.45 & 	 0.28 &  0.38 & 4.6	&  21:43:00.81 &  54:56:36.15 &  17.02 &   1.03  &	   &  17.1 \\
      21:43:00.56 &  54:56:23.14 &   17.84 & 	  0.4 &  0.85 & 6.6	&  21:43:00.86 &  54:56:14.13 &  17.62 &    1.07  &	   &  17.7 \\
      21:43:00.64 &  54:56:37.60 &   18.32 & 	 0.85 &  1.20 &  14.0 (3)  &  21:43:00.99 &  54:56:21.54 &  18.04 &   0.96  &	   &  15.9 \\
      21:43:00.65 &  54:56:26.97 &   14.97 & 	 0.26 &  0.50 & 4.3	&  21:43:01.01 &  54:56:24.08 &   17.75 &   1.03  &	   & 17.0 \\
      21:43:00.69 &  54:56:55.68 &   16.83 & 	 0.62 &  1.40 &  10.2	&  21:43:01.02 &  54:56:15.82 &  17.40 &   1.38  &	   & 22.7 \\
      21:43:00.70 &  54:56:20.44 &   17.78 & 	 0.58 &  0.77 & 9.6	&  21:43:01.06 &  54:56:25.42 &  18.12 &   1.07  &	   & 17.7 \\
      21:43:00.76 &  54:56:49.47 &   17.19 & 	 0.17 &  0.99 & 2.8	&  21:43:01.21 &  54:56:26.32 &  18.33 &   1.09  &	   & 18.0 \\
      21:43:00.79 &  54:56:35.64 &   15.27 & 	 0.92 &  1.32 &  15.2 (4)  &  21:43:01.24 &  54:56:22.26 &  17.14 &   1.51  &	   &  25.0 \\
      21:43:00.83 &  54:55:57.45 &   16.14 & 	 0.28 &  0.60 & 4.6	&  21:43:01.26 &  54:56:24.54 &  18.17 &   1.43  &	   &  23.5 \\
      21:43:00.83 &  54:56:01.76 &   13.96 & 	 0.29 &  0.56 & 4.8	&  21:43:01.30 &  54:56:24.11 &  17.53 &   1.22  &	   & 20.1 \\
      21:43:00.85 &  54:56:18.51 &   13.04 & 	 0.42 &  0.88 & 6.9	&  21:43:01.34 &  54:56:16.00 &  17.46 &   1.80  &	   & 29.8 \\
      21:43:00.90 &  54:56:47.41 &   16.97 & 	  0.2 &  0.96 &   3.3	&  21:43:01.36 &  54:56:41.54 &  17.73 &   0.99  &	   & 16.5 \\
      21:43:00.95 &  54:56:06.00 &   16.02 & 	 0.28 &  0.88 & 4.6	&  21:43:01.36 &  54:56:12.20 &  17.99 &   2.16  &	   & 35.6\\
      21:43:00.99 &  54:56:16.75 &   16.98 & 	 1.41 &  1.65 &  23.3 (5)  &  21:43:01.46 &  54:56:18.17 &  14.71 &   4.26  &	   &  70.4 \\
      21:43:01.16 &  54:56:33.20 &   17.13 &	 0.75 &  1.60 & 12.4    &  21:43:01.50 &  54:56:27.07 &  16.95 &   1.94  &	   & 32.0 \\
      21:43:01.19 &  54:56:16.91 &   15.46 &	 0.45 &  0.50 & 7.4	&  21:43:01.56 &  54:56:14.64 &  17.78 &   1.14  &	 &   18.8 \\ 
      21:43:01.19 &  54:56:32.46 &   16.97 &	  0.7 &  1.07 & 11.6 (6)	&  21:43:01.63 &  54:56:15.95 &  18.66 &   2.60  &	   &  42.9 \\
      21:43:01.23 &  54:55:59.62 &   15.48 &	 0.41 &  0.90 & 6.8	&  21:43:01.69 &  54:56:28.08 &  18.52 &   1.38  &	   &  22.7 \\
      21:43:01.23 &  54:56:36.56 &   17.26 &	 0.39 &  0.70 & 6.4	&  21:43:01.69 &  54:56:10.80 &  17.95 &   1.42  &	   &  23.4 \\
      21:43:01.44 &  54:56:24.06 &   15.56 &	 1.71 &  2.29 & 28.2 (7)	&  21:43:01.80 &  54:56:21.16 &  18.20 &   3.52  &	   &   58.1 \\
      21:43:01.63 &  54:56:04.83 &   14.53 &	 0.23 &  0.39 & 3.8	&  21:43:01.84 &  54:56:18.33 &  18.18 &   2.21  &	   &  36.5 \\
      21:43:01.92 &  54:56:07.50 &   14.89 &	 0.26 &  0.56 & 4.3	&  21:43:01.87 &  54:56:17.41 &  18.18 &   1.39  &	   &  22.9 \\
      21:43:01.95 &  54:56:35.51 &   13.21 &	 0.63 &  0.88 & 10.4 (8)	&  21:43:01.91 &  54:56:39.18 &  18.43 &   0.95  &	   &   15.6 \\
      21:43:01.97 &  54:56:51.85 &   16.38 &	 0.44 &  0.72 & 7.3	&  21:43:01.92 &  54:56:20.31 &  17.53 &   1.64  &	   &   27.0 \\
      21:43:02.11 &  54:55:56.38 &   16.77 &	 0.64 &  1.05 & 10.6 (9)	&  21:43:01.94 &  54:56:28.33 &  17.25 &   1.49  &	   &   24.6 \\
      21:43:02.12 &  54:56:01.49 &   17.95 &	 0.48 &  0.62 & 7.9	&  21:43:02.07 &  54:56:39.90 &   18.23 &   0.95  &	   &   15.7 \\
      21:43:02.17 &  54:56:35.20 &    17.3 &	 0.79 &  1.09 & 13.0 (10)	&  21:43:02.27 &  54:56:20.27 &  18.13 &   1.24  &	   &  20.4 \\
      21:43:02.30 &  54:56:14.84 &   15.72 &	 0.62 &  0.99 & 10.2 (11)	&  21:43:02.27 &  54:56:03.19 &  18.97 &    1.54  &	   &	25.4 \\
      21:43:02.46 &  54:56:02.76 &   15.14 &	 0.24 &  0.48 & 3.9	&  21:43:02.65 &  54:56:30.76 &  17.96 &    1.36  &	   &	22.4 \\
      21:43:02.95 &  54:56:08.53 &   16.39 &	 0.27 &  0.64 & 4.5	&  21:43:02.70 &  54:56:36.25 &  18.03 &   1.05  &	   &   17.3 \\
      21:43:03.02 &  54:56:46.77 &   14.79 &	 0.28 &  0.56 & 4.6	&  21:43:03.43 &  54:56:36.83 &  18.72 &   1.31  &	   &  21.6 \\
      21:43:03.09 &  54:56:43.78 &   16.41 &	 0.68 &  1.24 & 11.2 (12)	&  21:43:03.60 &  54:56:12.62 &  18.45 &   1.15  &	   &  18.9 \\
\hline
\end{tabular}
\end{table*}

\begin{table*}
 \centering
  \caption{UKIRT JHK photometry of sources in the central cluster (Cluster 2C) of IRAS 21407+5441. 
  The numbers in brackets in the A$_V$ column identify stars with IR excesses in T Tauri region and beyond. 
  The stars listed after the gap are detected only in HK bands.}
\begin{tabular}{cccccc||cccccc}
  \hline
R.A.(2000) & Dec.(2000) & H  & [H-K]  & [J-H] & A$_V$ & R.A.(2000) & Dec.(2000) & H  & [H-K]  & [J-H] & A$_V$ \\
\hline
21:42:20.64 & 54:55:36.51 &   17.56  &    0.49 & 0.92  &  8.1   &  21:42:24.39 & 54:55:33.43 &   17.66  &    0.33 & 0.83  & 5.5  \\
21:42:20.85 & 54:54:37.49 &   17.92  &     0.4 & 1.06  &  6.6   &  21:42:24.40 & 54:55:09.29 &   17.01  &    0.69 & 1.38  & 11.4 \\
21:42:20.91 & 54:55:00.80 &   12.79  &     0.4 & 0.94  & 6.6	&  21:42:24.43 & 54:55:11.32 &   15.63  &    1.07 & 1.68  & 17.7 (11) \\
21:42:21.03 & 54:55:19.35 &   13.37  &    0.25 & 0.52  & 4.1	&  21:42:24.45 & 54:55:07.69 &   17.19  &    1.27 & 2.19  & 21.0 (12) \\
21:42:21.29 & 54:55:19.23 &   16.35  &    0.34 & 0.76  & 5.6	&  21:42:24.48 & 54:55:03.95 &   17.32  &    0.61 & 0.87  & 10.1 (13) \\
21:42:21.41 & 54:55:00.26 &   16.83  &    0.49 & 0.94  & 8.1	&  21:42:24.58 & 54:55:38.36 &   16.05  &    0.35 & 0.73  & 5.8  \\
21:42:21.53 & 54:55:37.20 &   17.33  &    0.37 & 1.05  & 6.1	&  21:42:24.79 & 54:54:44.70 &   14.58  &    0.21 & 0.52  & 3.5  \\
21:42:21.71 & 54:54:54.88 &   17.84  &    0.59 & 0.84  & 9.7 (1)	&  21:42:24.98 & 54:55:21.07 &    16.8  &    0.46 & 1.02  & 7.6  \\
21:42:21.79 & 54:54:49.37 &    14.9  &    0.27 & 0.63  & 4.5	&  21:42:25.20 & 54:54:47.25 &   16.17  &     0.3 & 0.76  & 5.0  \\
21:42:21.80 & 54:55:39.18 &   15.47  &    0.48 & 1.05  & 7.9	&  21:42:25.89 & 54:55:31.79 &    17.2  &    0.34 & 1.00  & 5.6  \\
21:42:22.04 & 54:55:12.50 &   17.67  &    0.36 & 1.02  & 5.9	&  21:42:25.98 & 54:54:36.65 &    16.4  &    0.26 & 0.89  & 4.3  \\
21:42:22.06 & 54:54:42.64 &   17.76  &    0.44 & 0.96  & 7.3	&  21:42:26.02 & 54:55:28.38 &   16.15  &    0.82 & 1.43  & 13.5 \\
21:42:22.08 & 54:55:29.92 &   15.96  &    0.35 & 0.68  & 5.8	&  21:42:26.03 & 54:54:37.83 &   15.65  &    0.34 & 0.67  & 5.6  \\
21:42:22.26 & 54:55:06.50 &   16.43  &    0.41 & 0.80  & 6.8 (2)	&  21:42:26.08 & 54:55:12.81 &   16.94  &    0.44 & 1.06  & 7.3  \\
21:42:22.55 & 54:55:30.27 &   17.75  &    0.41 & 1.02  & 6.8	&  21:42:26.33 & 54:54:48.86 &     18.  &    0.39 & 0.75  & 6.4  \\
21:42:22.98 & 54:54:39.73 &    16.8  &    0.42 & 0.93  & 6.9	&  21:42:26.55 & 54:55:31.71 &   16.33  &    0.62 & 1.17  & 10.2 \\
21:42:23.04 & 54:54:43.89 &   18.05  &    0.47 & 0.95  & 7.8	&  21:42:26.68 & 54:55:00.08 &   16.41  &    0.25 & 0.79  & 4.1  \\
21:42:23.24 & 54:55:07.17 &   16.28  &    1.38 & 1.68  & 22.8 (3)	&              &             &          &         &	  &      \\
21:42:23.25 & 54:55:36.96 &   15.74  &    1.46 & 1.72  & 24.1	&              &             &          &         &	  &      \\
21:42:23.38 & 54:55:06.46 &   14.47  &    0.35 & 0.65  & 5.8 (4)	&  21:42:22.31 & 54:55:16.20 &   17.257 &   2.42  & 	  & 39.9 \\
21:42:23.42 & 54:55:05.13 &   16.68  &    0.74 & 1.27  & 12.2 (5)	&  21:42:22.44 & 54:55:03.73 &	 18.475 &   1.79  &	  & 29.5 \\
21:42:23.50 & 54:55:04.44 &   16.65  &    0.86 & 1.45  & 14.2 (6)	&  21:42:23.01 & 54:55:13.73 &	 17.839 &   1.25  &	  & 20.6 \\
21:42:23.59 & 54:54:42.54 &   13.68  &    0.47 & 1.10  & 7.8	&  21:42:23.56 & 54:55:24.01 &	 18.077 &   0.97  &	  & 16.0 \\
21:42:23.64 & 54:55:05.24 &   12.96  &    0.48 & 0.80  & 7.9 (7)	&  21:42:23.64 & 54:55:06.78 &	 18.215 &   1.09  &	  & 18.0 \\
21:42:23.72 & 54:54:58.88 &   16.69  &    0.64 & 1.47  & 10.6	&  21:42:23.67 & 54:55:15.34 &	 18.228 &   1.27  &	  & 21.0 \\
21:42:23.74 & 54:54:47.99 &   16.84  &    0.44 & 0.94  & 7.3	&  21:42:23.76 & 54:55:13.58 &	 16.511 &   1.27  &	  & 21.0 \\
21:42:23.83 & 54:55:07.48 &   15.81  &    1.25 & 1.80  & 20.6 (8)	&  21:42:24.50 & 54:54:59.86 &	 17.341 &   1.55  &	  & 25.6 \\
21:42:23.84 & 54:55:04.24 &	17.  &    0.84 & 1.78  & 13.9	&  21:42:24.57 & 54:55:08.08 &	 17.566 &   1.44  &	  & 23.8 \\
21:42:23.94 & 54:55:38.03 &   18.13  &    0.54 & 1.07  & 8.9	&  21:42:24.65 & 54:55:08.02 &	 17.335 &   0.97  &	  & 16.0 \\
21:42:23.96 & 54:55:05.49 &   16.73  &    1.04 & 1.59  & 17.2 (9)	&  21:42:25.43 & 54:55:20.59 &	 17.449 &   0.97  &	  & 16.0 \\
21:42:24.07 & 54:54:40.22 &   15.26  &    0.26 & 0.59  & 4.3	&  21:42:26.31 & 54:55:38.50 &	 18.324 &   1.36  &	  & 22.4 \\
21:42:24.14 & 54:55:03.66 &   14.33  &    1.02 & 1.22  & 16.8 (10)	&  21:42:26.34 & 54:55:35.41 &	 18.653 &   1.42  &	  & 23.4 \\
\hline
\end{tabular}
\end{table*}

\begin{table*}
 \centering
  \caption{UKIRT JHK photometry of sources in the north-eastern cluster (Cluster 2N) of IRAS 21407+5441. 
  The numbers in brackets in the A$_V$ column identify stars with IR excesses in T Tauri region and beyond.
  The stars listed after the gap are detected only in HK bands.}
 \begin{tabular}{cccccc||cccccc}
  \hline
R.A.(2000) & Dec.(2000) & H  & [H-K]  & [J-H] & A$_V$ & R.A.(2000) & Dec.(2000) & H  & [H-K]  & [J-H] & A$_V$ \\
\hline
 21:42:25.13 & 54:55:48.62 &   17.52 &    1.01 & 1.61 &  16.7 (1) &  21:42:28.22 & 54:56:00.03 &	11.98 &    0.32 & 0.9  & 5.3
 (10) \\
 21:42:25.89 & 54:55:31.79 &	17.2 &    0.34 & 1.00 & 5.6    &  21:42:28.43 & 54:55:58.35 &	15.29 &    0.67 & 1.00 & 11.1 (11)\\
 21:42:25.91 & 54:55:56.87 &   17.95 &    0.22 & 1.17 & 3.6    &  21:42:28.50 & 54:56:00.07 &	15.33 &    0.62 & 1.02 & 10.2 (12)\\
 21:42:26.55 & 54:55:31.71 &   16.33 &    0.62 & 1.17 & 10.2 (2)  &  21:42:28.51 & 54:55:49.85 &	14.32 &     0.2 & 0.68 & 3.3  \\
 21:42:26.82 & 54:55:53.67 &   16.29 &    1.25 & 2.37 & 20.6 (3)  &  21:42:28.67 & 54:55:56.46 &	14.17 &    0.37 & 0.81 & 6.1  \\
 21:42:26.94 & 54:55:45.79 &   12.48 &     0.5 & 1.13 & 8.3    &  21:42:28.92 & 54:55:34.90 &	 16.5 &    0.24 & 0.92 & 4.0  \\
 21:42:27.22 & 54:55:42.23 &   17.01 &    1.02 & 2.00 & 16.8 (4)  &  21:42:29.36 & 54:55:40.64 &	18.29 &     0.2 & 0.95 & 3.3  \\
 21:42:27.42 & 54:55:56.12 &   14.54 &    0.44 & 0.93 & 7.3    &              &             &         &         &      &      \\
 21:42:27.50 & 54:56:01.40 &   16.03 &    1.14 & 2.04 & 18.8 (5)  &              &             &         &         &      &       \\
 21:42:27.58 & 54:55:53.90 &   15.48 &    1.32 & 1.65 & 21.8 (6)  &  21:42:25.73 & 54:55:48.19 &   18.10 &  0.98   &      & 16.2  \\
 21:42:27.60 & 54:55:47.69 &   15.00 &    0.58 & 0.99 & 9.6  (7)  &  21:42:26.31 & 54:55:38.50 &	18.32 &  1.36   &      & 22.4  \\
 21:42:27.72 & 54:56:02.33 &   15.11 &    0.61 & 1.35 & 10.1   &  21:42:26.34 & 54:55:35.41 &	18.65 &  1.42   &      & 23.4  \\
 21:42:27.76 & 54:55:54.96 &   15.82 &    0.61 & 1.21 & 10.1 (8)  &  21:42:26.54 & 54:55:55.66 &	18.06 &  1.11   &      & 18.3  \\
 21:42:27.96 & 54:55:50.86 &   14.88 &    0.59 & 1.02 & 9.7  (9)  &  21:42:26.87 & 54:55:38.48 &	17.86 &  1.34   &      & 22.1  \\
 21:42:28.18 & 54:55:37.95 &	13.3 &    0.35 & 0.95 & 5.8    &  21:42:27.38 & 54:55:45.87 &	17.02 &  0.97   &      & 16.0  \\
 21:42:28.20 & 54:55:57.99 &   14.84 &    0.93 & 1.08 & 15.3   &  21:42:27.39 & 54:55:53.07 &	18.18 &  1.26   &      & 20.8  \\
\hline
\end{tabular}
\end{table*}

\section{Discussion}

\subsection{Stellar content in the three clusters}

In the case of Cluster 1 (Fig 2), we find a few YSOs close to the 
bright UCHII region at the centre. Notable among these are the star no. 7 
(Fig 5, top panel) which is of spectral type earlier than O8 and  
the star no. 5 which is a B2 type star. Also present closely surrounding 
the bright central UCHII region are several sources that are detected 
only in HK bands (but not in J band to the detection limit). 
Notable among these "red objects" is the one north of the star 7
(marked with a circle around it in Fig 2 and by a filled square 
in the CM-D in Fig 5, top right panel) having a [H-K] color 
of nearly 2 and of spectral type earlier than B0. 
Interestingly its position very nearly 
coincides with that of the centre of the CO
outflows detected in this object by Shepherd \& Churchwell
\cite{shep96}. We notice a number of PMS stars
very closely surrounding the UCHII region: the formation of which was 
possibly triggered by the expanding shock front.  
It is to be noted also that the bright 
central core of the UCHII region is detected only in HK bands 
with a [H-K] color of more than 4 (see Table 1). The CHII region towards
the north appears to be powered by a O9 type star (no. 8) that shows
moderate excess as compared to the stars in UCHII region towards its
south. The stars 5 and 7 suffer large visual extinction 
(as seen in the C-MD in Fig 5), A$_V$, of 23.3 and 28.2 mag (Table 1) and 
may represent PMS stars of Class I. As mentioned earlier, the spectral classes
assigned purely on the basis of C-MDs, could have been over-estimated, 
especially for the objects with large color excesses. 

In Cluster 2C (Fig 3), a compact HII region 
can be very clearly seen in the K-band
image of UKIRT encompassing the cluster, with a number of curved filamentary structures. 
In Fig 3 one can notice that the CHII region has a sharp edge towards west. 
The presence of stars earlier than B3 prompts us to believe that 
the faint halo around the cluster is a CHII region. Our radio 
continuum observations show a very intense 
emission surrounding the cluster suggesting a HII region (see Figs 6 and 7).  
There are about 13 stars in this cluster which fall in T Tauri 
region (Fig 5, middle panel). We find at least 11 stars that are between
spectral types earlier than B5 (Fig 5, middle panel and Table 2). 
For a few early type stars the A$_V$ is quite high (as seen in the C-MD in Fig 5 middle right panel),: 
star no. 8, a B0.5 type with 20.6 mag; star no. 11, B1 type with 17.7 mag; 
star no. 3, a B1 type with 22.8 mag which may be a Class I PMS star, star no. 12 of B2.5 type 
with 21.0 mag and star no. 9 of B3 type with an A$_V$ of 17.2 mag. 
The stars 7 and 10 having spectral type
of O9-O9.5 with A$_V$ 7.9 and 16.8 mag respectively, 
seem to be the main sources for the HII region. 
The star 4 has A$_V$ of nearly 5.8 mag 
showing that it may be of a ZAMS star of spectral type B2.    

At an angular separation of about 90$^{\prime\prime}$, to the north-east of  
the central cluster of IRAS 21407+5441, lies another cluster, Cluster 2N (Fig 4). 
Our UKIRT images cover it nearly completely (as verified from the 2MASS 
image of the region). There are about 12 stars in this cluster that fall into 
the T Tauri zone (Fig 5, bottom panel). 
At the centre of the cluster is a O8 ZAMS star (star 10 in UKIRT image in Fig 4) 
that showed an A$_V$ of 5.3 (as seen in the C-MD in Fig 5 bottom right panel and listed in Table 3). 
Star 6 (O9.5 type) and star 3 (B1 type) suffer large A$_V$ of 21.8 and 20.6 mags
respectively, with the former likely to be a Class I PMS star; 
while star 5 (type B1) and star 4 (type B3) 
suffer 18.8 and 16.8 mags respectively. 

In all the three clusters discussed above the stars fall in to or a little 
beyond the T Tauri zone (see Fig 5) and have (H-K) colors more than 1.0. 
These sources may be considered to be mostly 
Class II with large NIR excess (Classical T Tauri stars) or extremely reddened 
early type ZAMS stars, having excess emission in K-band 
(e.g., Blum, Damineli \& Conti \cite{blum01} and references therein). 
Our CM-Ds support the latter
scenario as also supported by the presence of HII regions.   
 
In addition to these three clusters, we looked into the 2MASS data on 
the region between the two IRAS sources. Interesting among the sources 
in this region is the Star A (marked in Fig 1 as well in Fig 5 for 
comparison with other stars). The 2MASS photometry shows 
that this object is a high mass star of spectral type earlier than O4 
with large color excess (J-H = 2.24 and H-K = 1.94) and hence 
suffering a visual extinction of 32 mags. 
It is possible that this star could belong to the molecular cloud from 
which the three clusters formed later. Alternatively 
this may be a background star, not associated with the cloud but 
obscured by it in the line-of-sight.    

The MSX A band contours shown in Fig 1 (and the image in Fig 6) indicate  
extended low-level ($\sim$ 2-4 $\times$ 10$^{-6}$ W/m$^{2}$/Sr) mid-infrared emission 
towards the south of Star A, apart from the more intense emission around the two IRAS sources.
Of the 9 MSX point sources present in the region (with detection quality 4 in the 8.2 $\mu$m A band), 
we find 5 are associated with the two IRAS sources (two in 
each) and Star A. It may be mentioned that the MSX fluxes correspond to several 
resolved sources in UKIRT images in each of the three clusters.    
The other 4 MSX point sources are present towards SE and SW of Star A. 
Most of these sources have fluxes good only in A band. 
We have extracted 2MASS archival JHK$_{s}$ data (of quality "A" in all the three bands) in this region and
found out that the Star A stands out with its large color excesses. 
The other stars (including the counterparts for the 4 MSX point sources), 
mostly falling between the two extinction vectors 
for main-sequence; and if assumed to be at the same distance as the two IRAS sources, 
represent massive stars of late O type or early B type. Hence we believe that they are 
foreground stars. Using the code DUSTY (Ivezic, Nenkova \& Elitzur \cite{ivez99}), 
we have modelled the SED of Star A, 
taking data from 2MASS 
and MSX archives and found that the SED fits well with a T$_{eff}$ $\sim$ 30000 K 
with emission also from graphite dust grains at T = 650 K, indicating 
that the star may be very massive and obscured. Interestingly Star A does not have 
a counterpart in IRAS data. If the Star A is indeed a massive PMS star with a UCHII around, then 
the emission should peak around 100 $\mu$m (e.g., Churchwell \cite{chur02}). So it appears that 
it is well beyond its PMS phase with warm dust/gas in an accretion disk. 

 \begin{figure*}
   \centering
\includegraphics[scale=0.6,angle=-0]{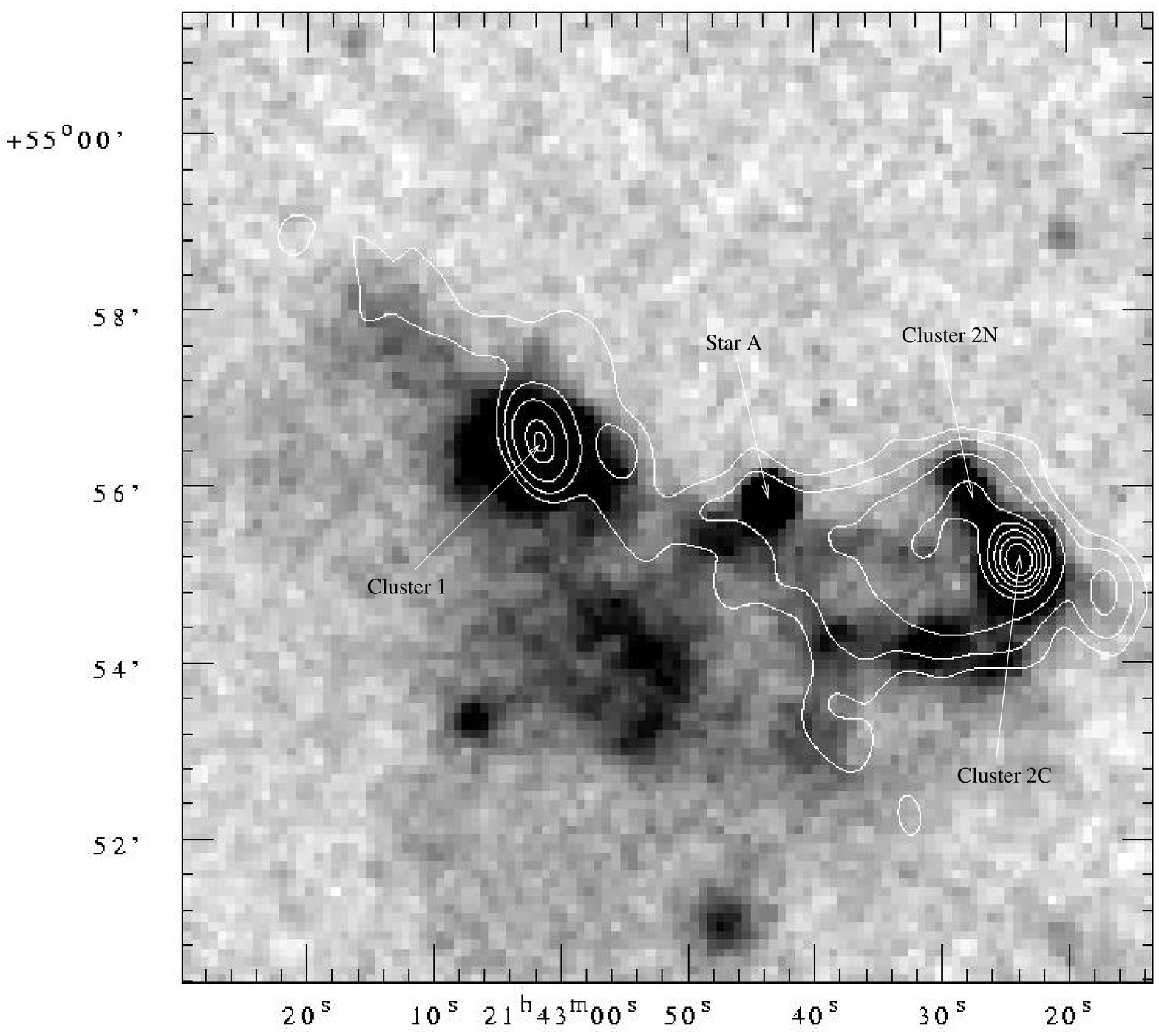}
   \caption{MSX A band (8.2 $\mu$m) low-resolution mosaic image of the region shown overlaid 
   by the 1.28 GHz radio continuum contours: Clusters 1, 2C and 2N are shown by arrows 
   as is the Star A (cf. Fig 1). The radio contours represent flux density values 
   of 10, 20, 40, 80, 100, 120, 140, 160 and 180 mJy/beam. The beam size is 30 arc sec.
   The abscissa (RA) and ordinate (Dec) are for J2000 epoch.}
              \label{msxradio}%
    \end{figure*}

There are about 43 stars with H-K color $\geq$ 0.50 in Cluster 1; about 30 in 
Cluster 2C and about 25 in Cluster 2N, as identified from UKIRT photometry. Assuming 
that they represent pre-main-sequence embedded sources, we can infer that the 
volume density for the three clusters seems to be less than that for a typical massive star 
forming cloud (e.g., Smith \cite{smit04}).  

\subsection{Parameters from Radio continuum}

Here in this section 
we derive some physical parameters from the radio fluxes and also 
estimate the distances to the sources by combining the radio and 
NIR data. 
Assuming that the
continuum emission is caused by free-free radiation from
accelerated electrons (thermal Bremsstrahlung), we expect a
power law for the spectral energy distribution with an index of
$\sim$ 0.1. However, Miralles et al \cite{mira94} have shown that
among several UCHII regions that they studied in two radio
frequencies (4.9 and 14.9 GHz), only the southern UCHII region
(in the Object 1) showed an index of 0.99 $\pm$ 0.01. It was
suggested by these authors that the object must be obscured by
dust.


  \begin{figure*}
   \centering
\includegraphics[scale=0.6]{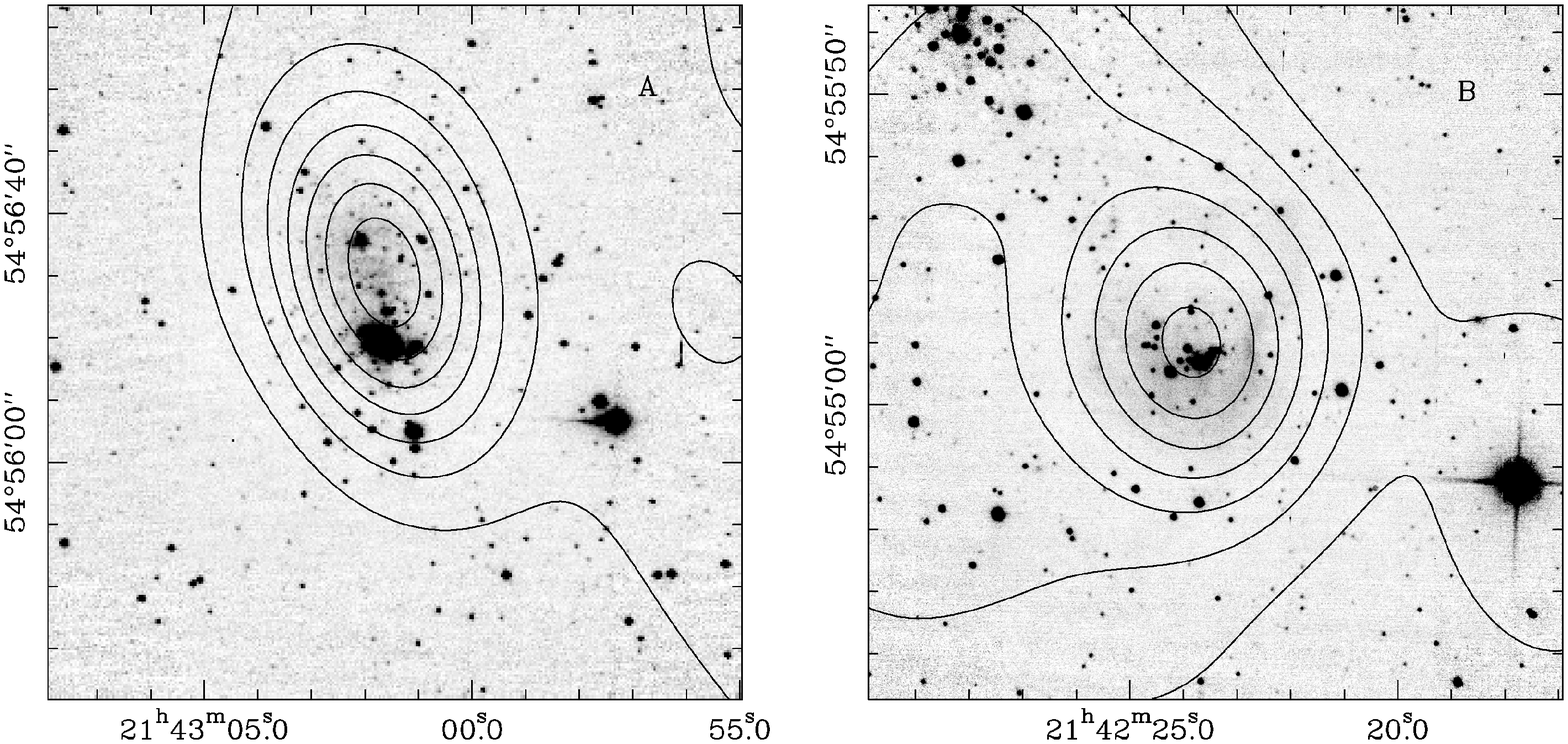}
   \caption{GMRT 1.28 GHz contours overlaid on the UKIRT K-band images of 
   IRAS 21413+5442 (A on the left) and IRAS 21407+5441 (B on the right). 
   The contour flux density levels correspond to 10-100 mJy/beam in steps 
   of 15 mJy/beam for A and 
   25-175 mJy/beam in steps of 25 mJy/beam. The beam size is 30 arc sec.
   The abscissa (RA) and ordinate (Dec) are for J2000 epoch.}
              \label{gmrtolob1}%
    \end{figure*}

              \label{gmrtolob2}%

Firstly, we try to determine the individual 1.28 GHz continuum
flux for the two HII regions in Object 1. Adopting a spectral index 
of $\sim$ 1.0 (from Miralles et al \cite{mira94}), one would expect a flux
of $\sim$ 30 mJy for the GMRT observation at 1.28 GHz. The
index for the northern HII region is 0.04 $\pm$ 0.07 from
Miralles et al \cite{mira94} observations. This would give a flux of $\sim$
50 mJy for the northern CHII region. Hence the 1.28 GHz
combined flux would be $\sim$ 80 mJy. The GMRT flux at 1.28
GHz is 103 mJy in a beam size of 30 arcsec that encompasses
both the HII regions. This is a reasonably fair comparison
considering various uncertainties involved. Partitioning the
flux in the same ratio, we estimate the flux due to the UCHII
region at 1.28 GHz to be $\sim$ 38 mJy and 65 mJy for CHII
region.

Comeron \& Torra \cite{come01} showed for an ionization bounded HII
region optically thin at frequency $\nu$, with an electron
temperature of T$_e \sim 10^4 K$, the distance may be given by

\begin{equation}
D(kpc) \approx  5.35 \times 10^{-3} [\frac{\nu^{0.1}(GHz)
S_{\nu}(J_y)}{\Sigma^{N}_{i=1}10^{-1.26 H_{oi}}}]^{0.233}
\end{equation}

where S$_\nu$ is the flux density in Jy at frequency $\nu$ (in
GHz). H$_{oi}$ is the de-reddened H-band magnitude; H$_o$ = H
(observed) - A$_H$, with A$_H$, the extinction in the H band.
A$_H$ may be estimated from the formula A$_H$ = 0.11 +
2.87(H-K), assuming the extinction law due to Rieke \& Labofsky
\cite{riek85}. The term 0.11 is to take care of the intrinsic
color for OB stars. Using this approximate formula, the GMRT
radio fluxes and the combined H band magnitudes of all the
early B type stars in the two HII regions in the Object 1 (Cluster 1), we
obtain an average value of 4.8 kpc; and for the Object 2 (Cluster 2C), we 
get an average value of 7.3 kpc. The kinematic distance of 7.4 kpc
was estimated by Wouterloot \& Brand \cite{wout89} and adopted by 
Shepherd \& Churchwell \cite{shep96}. However, the distance
problem that arises due to excess photon flux (if placed
nearer) may be circumvented due to the presence of more than
one ionizing photon source. Our distance estimates are in good 
agreement with the value of 7.4 kpc adopted by earlier workers.
 
Using the GMRT fluxes we then estimated the physical parameters
such as electron density, emission measure and optical depth, following 
the standard work of Mezger \& Henderson \cite{mezg67}, under the 
assumption of spherical HII region. Assuming the
angular sizes of the two HII regions (UC HII region $\sim$ 1.6 arcsec
and CHII region $\sim$ 12 arc sec) from Miralles et al.
\cite{mira94}, and a typical electron temperature of 10$^4$
K, we get the elctron density N$_e$ = 8.9 $\times$ 10$^3$ cm$^{-3}$ 
and the emission measure EM = 6.6 $\times$ 10$^6$
cm$^{-6}$ pc for the UCHII region and N$_e$ = 5.7 $\times$ 10$^2$ cm$^{-3}$
and EM = 2.0 $\times$
10$^{5}$ cm$^{-6}$ pc for the CHII region. If we adopted lower
T$_e$ (3300 K as derived by Miralles et al \cite{mira94}), we get values of
EM about 68\% of the values with T$_e = 10^4$ K. 

Having obtained the EM values, we can now calculate the optical
depth values. For the GMRT fluxes, assuming T$_e = 10^4$ K, we
get $\tau_c$ (1.28 GHz) = 1.29 for the UCHII region and 0.04
for the CHII region. If we adopt T$_e = 3500 K$ (see Miralles
et al 1994), we get $\tau_c = 5.67$ for the UCHII region and
0.17 for the CHII region. This shows that the UCHII region is
optically thick at 1.28 GHz (compare Miralles et al \cite{mira94} for 4.9 and
14.9 GHz). This suggests that the HII region is not ionization bounded 
and hence the eqn. 1 may not be adequate for its distance estimation (for more 
discussion on this point, see Comeron \& Torra \cite{come01}).  

Similar calculations were performed for Object 2 (central cluster) also. We
obtained a value of N$_e$ = 2.8 $\times$ 10$^2$ cm$^{-3}$ and EM $ = 1.2 \times 10^5$ cm$^{-6}$ pc,
assuming the angular size of the source to be $\sim$ 30 arcsec.
This yields the continuum (1.28 GHz) optical depth $\tau_c =
0.023$ showing it is optically thin at this frequency. This could be the reason 
why this object was not detected in CS(2-1) line survey of Bronfman et al. \cite{bron96}, 
since the line gets preferentially populated in denser regions. From the parameters 
derived from the radio continuum it may be inferred that while the UCHII region is obviously  
the youngest, the CHII region in Cluster 1 is younger than the CHII region in Cluster 2C which 
in turn could be younger than Cluster 2N.

We then estimated the Lyman continuum photon fluxes from 
the N$_e$ values and sizes of the HII regions (e.g., Smith \cite{smit04}): 
2.3 $\times$ 10$^{47}$ s$^{-1}$ for the UCHII region and 7.8 $\times$ 10$^{47}$ s$^{-1}$ for the 
CHII region in Object 1 and 2.9 $\times$ 10$^{48}$ s$^{-1}$ for the CHII region in Cluster 2C (Object 2).
If we consider the Lyman continuum fluxes for ZAMS stars of same spectral types as given in 
Tables 1-3, we obtain for Cluster 1 (adding the UCHII and CHII), a total of 4.6 $\times$ 10$^{48}$ s$^{-1}$ 
and for Cluster 2C a total of 1.8 $\times$ 10$^{48}$ s$^{-1}$ (from Thompson \cite{thom84}).  
Our results are consistent with multiple early B type 
or late O type stars that account for the photon fluxes. 
Thus, within uncertainties, the derived radio 
properties support the NIR photometric results.  

A similar analysis of radio continuum emission around Star A suggests that 
the emission region is optically thin ($\tau_c = 0.006$) and the photon flux is 
consistent with a late O type star.  

\subsection{Association of the three clusters and the presence of Star A}

It is clear from our infrared photometric study as well as the previous 
radio studies that the two IRAS sources are most likely associated with each other,  
probably belonging to the same parent molecular cloud. One can notice in Fig 5 that the 
colors of YSOs in the three clusters are nearly the same, although those in Cluster 1
are slightly redder in comparison. Age estimates (Palla \& Stahler \cite{pall00}) 
from the color-magnitude 
diagram of absolute J  vs (J-H) showed that all the three clusters have similar 
ages with all the moderately massive stars being close to ZAMS.  
It should be mentioned here that while the two IRAS sources are nearly at the 
same kinematic distance (from published radio observations of Wouterloot \& Brand \cite {wout89};  
Carral et al \cite{carr99}), attributing them to the same molecular cloud 
should reconcile with a large velocity gradient across the cloud 
(from the published CO line data of Yang et al \cite{yang02}).

What is more intriguing, however, is the presence 
of Star A between the two sources (at a projected separation of $\sim$ 10 pc from either,
assuming the same distance for the Clusters and Star A)
and the indication of its possible association with the three clusters as can be seen 
from the MSX A-band image (Fig 6). We shall examine the possibility for Star A 
to have initiated the star formation in the three clusters.

It was proposed and subsequently evidences were shown by several 
authors that young massive stars can trigger star formation in the surrounding 
interstellar medium, either by their expanding HII regions sweeping up material 
that may collapse by a gravitational perturbation (Elmegreen \& Lada \cite{elme77}) 
or by the radiation-driven collapse of already present clumpy matter 
(Lefloch, Lazareff \& Castets \cite{lefl97}, Karr \& Martin \cite{karr03}, 
Zinnecker \& Yorke \cite{zinn07} and the references therein). 

The enigmatic single star (Star A) located in the region between the two IRAS sources suggests 
that this star being very massive, its winds/radiation might have prevented 
any further star formation in the near vicinity. 
But it is possible that farther away, the winds/radiation originated from this star might have 
triggered the intermediate to moderately high mass star formation in the three clusters 
flanking it. But we need to reconcile with the large extinction that the star suffers.

The large extinction may be attributed to an accretion disk.  
Being a massive star, it is possible that the winds and energetic radiation are 
switched on (upon entering the ZAMS stage), even before accretion phase ceases. It is also possible 
that the winds and radiation from the star could escape through the polar regions and interact 
with the surrounding matter. The formation time scale for a massive star  
is about 5 $\times$ 10$^5$ yrs; the times scales for the accretion disk to 
evaporate following the onset of energetic radiation and winds from the 
pre-main-sequence star are also of the same order (see e.g., Churchwell \cite{chur02}; 
McKee \& Ostriker \cite{mcke07}.

However, the non-detection of Star A in the far-infrared 
(beyond 25 $\mu$m) by IRAS (or COBE) casts doubts 
on being massive as argued by us and hence on the triggering scenario. 
Alternatively then, the Star A could be a nearby foreground star 
in which case its spectral type would be 
that of an intermediate-low mass pre-main-sequence star suffering high extinction. 

Further deeper observations are required on the entire 
molecular cloud complex using an appropriate molecular tracer in order to 
resolve both these issues, viz. the large velocity gradient across the cloud and 
the nature of Star A.

\section{Conclusions}

The important conclusions of this work are:

(i) Seeing limited near-infrared photometry was made on two massive 
star forming regions IRAS 21413+5442 and IRAS 21407+5441 using UKIRT facility. 
A radio continuum mapping of the two regions was also done using GMRT facility
in the 1.28 GHz band. 

(ii) Several embedded early B type stars are 
detected in both IRAS 21413+5442 (Cluster 1) and IRAS 21407+5441 (Clusters 2C and 2N).
Several pre-main-sequence stars are detected only in H and K bands 
having large (H-K) colors (red objects) in all the three clusters. 
One such red object in Cluster 1, possibly
a late O type star is located very close to the centre of the CO outflows in IRAS
21413+5442. A new CHII region was identified in the Cluster 2C which is
powered by early B type or late O type stars. 

(iii) Our infrared photometry and the radio map along with the MSX A band image 
indicate that the two IRAS sources are associated with each other.
 A highly obscured massive (O4 or earlier) star is 
shown to be present in the region between the two IRAS sources. 
It is conjectured that this single massive star 
might have triggered the star formation in the clusters. Further investigation
is needed to establish this hypothesis beyond doubt.  

(iv) The radio continuum observation at 1.28 GHz confirms the
earlier results that the UCHII region in IRAS 21413+5442 is
optically thick with the spectral index 1.0. The radio properties 
seem to support the NIR photometric results. \\

\section*{acknowledgements}
   We thank the anonymous referee for a stimulating report that helped  
   in improving the original version of the manuscript. 
   This research was supported by the Department of Space,
   Government of India. We
   thank the staff of GMRT (NCRA-TIFR) at Khodad, Central India for
   their help
   in obtaining the radio observations. We also thank the staff
   of UKIRT, Hawaii for
   observational time and help.  
   This publication makes use of data products from the 2MASS, 
   which is a joint project of the University of Massachusetts 
   and the IPAC/California Institute of 
   Technology, funded by NASA and NSF. 
   This research has also made use of the data products from the MSX, 
   NASA/IPAC Infrared Science Archive, which is operated by the 
   Jet Propulsion Laboratory, California Institute of Technology, 
   under contract with NASA. The use of DUSTY code is thankfully  
   acknowledged.

\end{document}